\documentclass[twocolumn,showpacs,secnumarabic,amssymb, nobibnotes, aps,
prb]{revtex4}

\usepackage{graphicx}
\usepackage{dcolumn}
\usepackage{bm}

\begin{document}


\title{Transformation of in-plane $\rho (T)$ in $YBa_{2}Cu_{3}O_{7-\delta}$ at fixed oxygen content}

\author{M. M. Abdelhadi}
\email{mabdel@phys.ualberta.ca}
 \affiliation{Department of Physics, University of Alberta, Edmonton, Alberta Canada T6G 2J1.}
\author{J.A. Jung}
 \email{jung@phys.ualberta.ca}
\affiliation{Department of Physics, University of Alberta,
Edmonton, Alberta Canada T6G 2J1.}

\date{\today}

\begin{abstract}
This paper reveals the origin of variation in the magnitude and
temperature dependence of the normal state resistivity frequently
observed in different YBCO single crystal or thin film samples
with the same $T_{c}$. We investigated temperature dependence of
resistivity in $YBa_{2}Cu_{3}O_{7-\delta}$ thin films with 7-
$\delta = 6.95$ and 6.90, which were subjected to annealing in
argon at 400-420 K ($120-140^{o}C$). Before annealing these films
exhibited a non-linear $\rho_{ab}(T)$, with a flattening below 230
K, similar to $\rho_{b}(T)$ and $\rho_{ab}(T)$ observed in
untwinned and twinned YBCO crystals, respectively. For all films
the annealing causes an increase of resistivity and a
transformation of $\rho_{ab}(T)$ from a non-linear dependence
towards a more linear one (less flattening). In films with 7-
$\delta = 6.90$ the increase of resistivity is also associated
with an increase in $T_{c}$. We proposed the model that provides
an explanation of these phenomena in terms of thermally activated
redistribution of residual O(5) oxygens in the chain-layer of
YBCO. Good agreement between the experimental data for
$\rho_{ab}(t,T)$, where t is the annealing time, and numerical
calculations was obtained.
\end{abstract}


\maketitle

\section{INTRODUCTION}
Right after discovery of high temperature superconductivity, it
has been observed that temperature dependence of the normal state
resistivity $\rho (T)$ is linear in many samples over a wide range
of temperatures. Phillips \cite{ref:Phillips} explained this
result (believed to be universal for all optimally doped HTSC)
using two-carrier percolation model. Later, the observation of a
linear $\rho (T)$ has been also used to support the argument that
non-Fermi liquid states are responsible for the normal state
properties of HTSC \cite{ref:Anderson}.

\begin{figure}
\def\picfilename{Fig.11n.eps}
\input epsf
\epsfxsize = 230pt \epsfysize =420pt \epsfbox{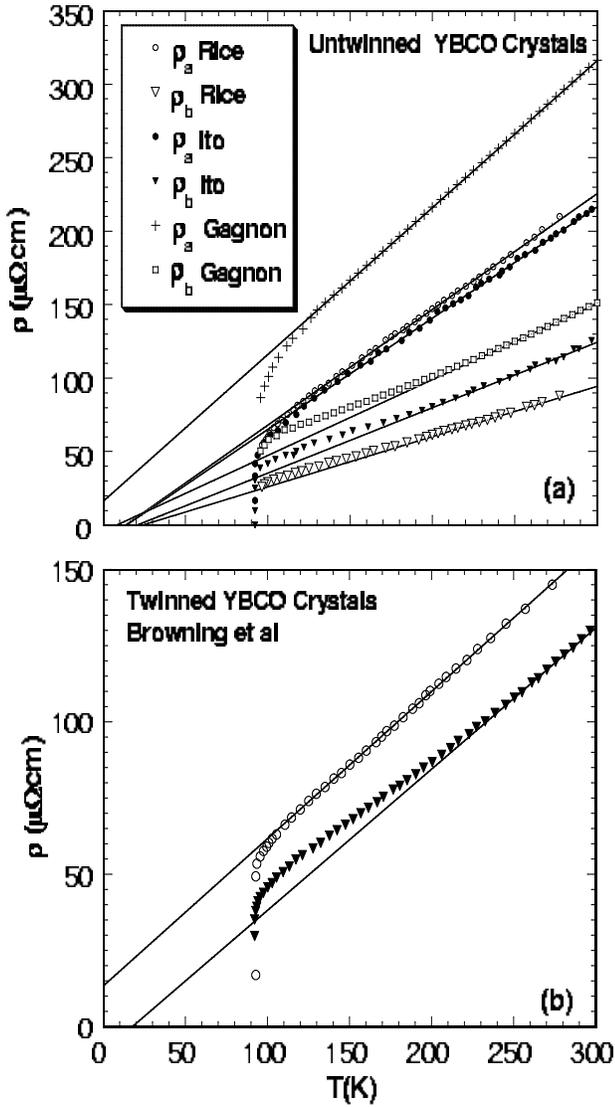}

\caption{\setlength{\baselineskip}{10pt} (a) Temperature
dependence of resistivities, $\rho_{b}(T)$ and $\rho_{a}(T)$,
measured in untwinned YBCO crystals along the chain direction
(b-direction) and along the a-direction perpendicular to the
chains, respectively. Note a flattening of $\rho_{b}(T)$ below 220
K and a linearity of $\rho_{a}(T)$ over a wide range of
temperatures for all three samples
\cite{ref:Rice,ref:Ito,ref:Gagnon}. (b)Temperature dependence of
the in-plane resistivity $\rho_{ab}(T)$ for twinned YBCO crystals.
Note that in this case $\rho_{ab}(T)$ exhibits either a non-linear
temperature dependence (with flattening) below 220 K or a linear
dependence in the sample with  higher resistivity
\cite{ref:Browning}.} \label{fig:fig1}
\end{figure}

The analysis of the experimental data for $\rho (T)$ obtained on
untwinned and twinned YBCO single crystals and YBCO thin films
revealed however that the pure linear temperature dependence of
resistivity is not the characteristic of $\rho (T)$ in HTSC. In
fact, it has been found that $\rho_{b}(T)$ measured along chain
direction in untwinned crystals of YBCO
\cite{ref:Friedmann,ref:Rice,ref:Ito,ref:Gagnon} exhibits a
non-linear behavior with a flattening  at temperatures below 220K
[see Fig.\ref{fig:fig1}(a)]. On the other hand, $\rho_{a}(T)$
measured in a direction perpendicular to the chains, increases
linearly with an increasing temperature over a wide temperature
range. The magnitude of $\rho_{a}(T)$ is about 2-2.5 times larger
than that of $\rho_{b}(T)$ at corresponding temperatures. An
exception is the result of Welp  et al \cite{ref:Welp}, who
observed a linear temperature dependence of both $\rho_{a}$ and
$\rho_{b}$. However, the absolute values of these resistivities
are about 2 times higher than those reported elsewhere for
untwinned YBCO crystals
\cite{ref:Friedmann,ref:Rice,ref:Ito,ref:Gagnon}.

The in-plane resistivity $\rho_{ab}(T)$ measured in twinned YBCO
crystals of $T_{c}=93K$, $\Delta T_{c}= 0.2K$, and of low
resistivity $\rho_{ab}(300K)=130 \mu \Omega cm $
\cite{ref:Browning} displays non-linear temperature dependence
similar to that observed in the most of untwinned YBCO crystals
along the chain direction [see Fig.\ref{fig:fig1}(b)].
Surprisingly, other YBCO twinned crystals of the same $T_{c}$ and
$\Delta T_{c}$, and higher room temperature resistivity
$\rho_{ab}(300K)=160 \mu \Omega cm$ \cite{ref:Browning} are
characterized by $\rho_{ab}$ with a linear dependence on
temperature [Fig.\ref{fig:fig1}(b)]. This value of resistivity is
close to that of a non-linear $\rho_{b}(T)$ measured at 300 K
along the chain-direction in some untwinned YBCO
crystals\cite{ref:Gagnon}.

\begin{figure}[ht]

\def\picfilename{Fig.22.eps}
\input epsf
\epsfxsize = 230pt \epsfysize =220pt \epsfbox{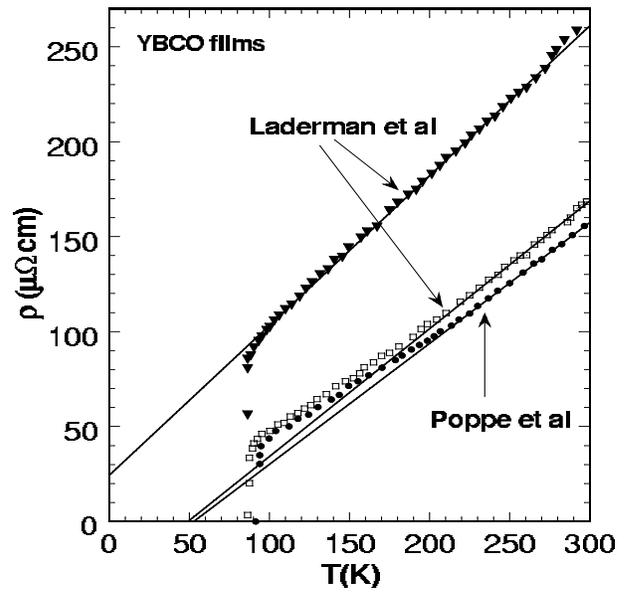}

\caption{\setlength{\baselineskip}{10pt} Temperature dependence of
the in-plane resistivity $\rho_{ab}$ measured in c-axis oriented
epitaxial YBCO thin films. The data indicate that films with low
resistivity exhibit a deviation from the linear temperature
dependence at temperatures below approximately 230K
\cite{ref:Poppe,ref:Laderman}.} \label{fig:fig2}
\end{figure}

Temperature dependence of $\rho_{ab}$ measured in c-axis oriented
epitaxial YBCO thin films shows features similar to those observed
in twinned YBCO crystals \cite{ref:Poppe,ref:Laderman}. Optimally
doped films with resistivity as low as $150 \mu \Omega cm$ at 300K
have non-linear $\rho_{ab}(T)$ with a flattening at temperatures
below 230K  \cite{ref:Poppe} (see Fig.\ref{fig:fig2}). However, it
appears that linear temperature dependence of $\rho_{ab}(T)$
characterizes films with higher resistivity (above approximately $
250 \mu \Omega cm$ at 300 K). Underdoped films with $T_{c} \simeq
84-85K$ exhibit $\rho_{ab}(T)$ similar to that observed in
optimally doped films \cite{ref:Laderman} (see
Fig.\ref{fig:fig2}).

Two different theoretical approaches were used so far to explain
the observed non-linear $\rho (T)$. The first one is based on
classical electron-phonon Bloch-Gr\"{u}neisen (BG) theory of
metallic conductivity \cite{ref:Poppe}. It predicts a non-linear
$\rho (T)$ for temperatures less than characteristic temperature
$T^{*}=\eta \Theta_{D}$ (where $\Theta_{D}$ is the Debye
temperature and $\eta \sim 1-2$), and a linear $\rho (T)$ for
temperatures above $T^{*}$. The other one employs a thermal
Frenkel disorder of long range-ordered chains of oxygen, a process
during which an oxygen atom jumps from its O(1) chain site to a
nearby O(5) interstitial (inter-chain) vacant site
\cite{ref:Goldschmidt}.

Poppe etal \cite{ref:Poppe} attempted to explain the non-linear
$\rho (T)$ observed in YBCO thin films by assuming that the
resistivity is the sum of the temperature independent residual
resistivity $\rho_{r}$ due to impurities or defects, and the
Bloch-Gr\"{u}neisen resistivity $\rho_{BG}(T)$. They found that BG
theory can be applied only if one assumes very high values of
Debye temperature within the range of 650-950K. Martin et
al\cite{ref:Martin} were able to obtain a good theoretical fit to
a non-linear $\rho (T)$ measured in YBCO(124) films at
temperatures up to 600K, using BG theory and an effective
transport Debye temperature $\Theta^{*}_{D}=(2k_{F}/G)\eta
\Theta_{D}$ of 500K. They argued that $\Theta^{*}_{D}$ can be
substantially smaller than $T^{*}=\eta\Theta_{D}$ if the Fermi
vector $k_{F}$ is sufficiently smaller than the reciprocal lattice
vector G.

Goldschmidt et al\cite{ref:Goldschmidt} interpreted the non-linear
temperature dependence of resistivity in fully oxygenated
polycrystalline YBCO in terms of a thermally activated process of
Frenkel pair formation, during which the chain oxygen jumps from
O(1) site to a nearby O(5) site. According to Goldschmidt et al
this process accounts for an excess electron scattering with an
increasing temperature. Each Frenkel pair creates an intrinsic
defect resistivity $\rho_{d}$ which represents an excess
resistivity connected in series to the perfect chain resistivity.
The upturn in resistivity that occurs at temperatures above 200K,
is explained as due to an increase of Frenkel pair density with an
increasing temperature. The excess resistivity is defined as:
$\rho_{excess}= \rho_{chain}- (\rho_{o}+ aT)= \rho_{d}C_{d}(T)$,
where $C_{d}(T)$ is the temperature dependent fractional defect
pair occupancy. Temperature dependence of $C_{d}$ was taken as
that for a two level system with the energy barrier $C_{d}\simeq
\exp(-E_{d}/kT)$,where $E_{d}$ is the energy barrier between the
final O(5) and initial O(1) oxygen sites. Consequently
$\rho_{excess}\simeq\rho_{d}\exp(-E_{d}/kT)$. Goldschmidt et al
obtained good theoretical fit to $\rho (T)$ data measured for a
granular YBCO taking the activation energy $E_{d}\simeq 120 meV$
and defect resistivity $\rho_{d}\simeq 2m \Omega cm$. Measurements
performed on untwinned YBCO crystals by Gagnon et al
\cite{ref:Gagnon} revealed however that temperature dependence of
the excess resistivity above 300K is linear, not at all
exponential.

The experimental results obtained on untwinned and twinned YBCO
crystals, and YBCO thin films could be summarized as follows:
Although $\rho (T)$ for the most of untwinned crystals show a
non-linear $\rho_{b}(T)$ and a linear $\rho_{a}(T)$, with the
magnitude of $\rho_{a}(T)$ being higher than that of
$\rho_{b}(T)$, these two forms of $\rho (T)$ have been also
randomly observed in twinned YBCO crystals and thin films.
Surprisingly, the absolute values of resistivity of twinned YBCO
crystals and films are often lower than those measured in some
untwinned YBCO crystals along the b-direction. Also, there is some
correlation between the magnitude of resistivity and a specific
form of $\rho (T)$ (non-linear or linear) in twinned YBCO crystals
and in YBCO films. In these cases the resistivities of a linear
$\rho (T)$ are in general higher than the corresponding ones for a
non-linear $\rho (T)$. In untwinned YBCO crystals, a non-linear
and a linear $\rho (T)$ has been attributed to the chain and plane
conductivities, respectively \cite{ref:Gagnon}. Therefore, the
random appearance of a non-linear and linear $\rho (T)$ in twinned
YBCO crystals and films is surprising and should be investigated.

In this paper we provide alternative explanation of the presence
of a non-linear and a linear dependence of resistivity $\rho_{ab}$
on temperature in YBCO. Our studies were focused on YBCO c-axis
oriented thin films, both optimally doped ($7-\delta\simeq 6.95$,
$T_{c}\simeq90-91K$) and slightly underdoped ($7-\delta\simeq
6.90$, $T_{c}\simeq 85K$), with a non-linear $\rho_{ab}(T)$. We
investigated the effect of oxygen ordering (redistribution) on
$\rho_{ab}(T)$ in those films. The applied experimental procedure
included systematic annealing of the samples in argon at 400-420K
($120-140^{o}C$), followed by the measurements of $\rho_{ab}(T)$
at temperatures between $T_{c}$ and 300K.

The results  revealed a transformation of $\rho_{ab}(T)$ from a
non-linear temperature dependence towards a more linear one (less
flattening) . This transformation is accompanied by an increase in
the absolute value of resistivity. In underdoped samples,
annealing causes also an increase in $T_{c}$. We present possible
explanation of this phenomenon, which is based on thermally
activated redistribution of oxygen in the chain-layer of YBCO.

\section{EXPERIMENTAL PROCEDURE}
We investigated c-axis oriented YBCO thin films, which were
deposited on different substrates: $SrTiO_{3}$,$LaAlO_{3}$ and
sapphire (with $CeO_{2}$ buffer layer) using the standard off-axis
rf magnetron and laser ablation techniques. The standard
photolithography  was used to produce $60\mu m$ wide and 6.4 mm
long thin film bridges with four measurement probes. The distance
between the voltage probes was 0.4 mm. Silver contacts were
sputter-deposited using rf-magnetron technique. The samples were
annealed in argon at temperatures $120-140^{o}C$ several times for
a total time between 8 to 24 hours. The measurements of
$\rho_{ab}(T)$ (in the a-b planes of the films) at temperatures
between $T_{c}$ and 300K, were conducted after each 2-5 hour
annealing period. During these measurements a dc current of $10
\mu A$ was applied to the sample in the form of short pulses (of
duration less than 200ms) in order to reduce Joule's heating.
Current reversal was used to eliminate the effects of thermal emf
in the leads. The voltage was measured using Keithley 2182
nanovoltmeter connected to Keithley 236 current source, with the
nanovoltmeter working as a triggering unit for the current source.
The nanovoltmeter operated in a "Delta" mode which allows one to
perform very fast multiple measurements of the voltage from two
voltage measurements for two opposite directions of the current.
Temperature was monitored by a carbon-glass thermometer and
stabilized (using the inductance-less heater and a temperature
controller) within $\pm 10mK$.

\section{EXPERIMENTAL RESULTS}
\begin{figure}[ht]
\def\picfilename{Fig.33.eps}
\input epsf
\epsfxsize = 230pt \epsfysize =420pt \epsfbox{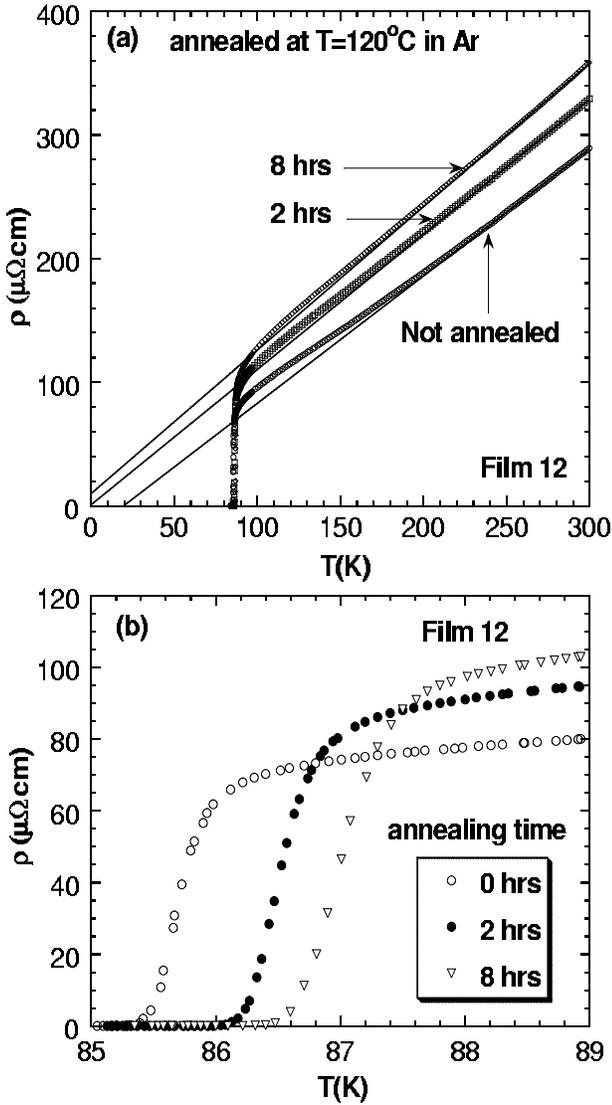}

\caption{\setlength{\baselineskip}{10pt}(a) Temperature dependence
of the in-plane resistivity $\rho_{ab}$ measured in a c-axis
oriented slightly underdoped YBCO thin film 12 with $T_{c}=85.4K$,
before and after annealing for 2 and 8 hours in argon at
$120^{o}C$. Note a systematic increase of resistivity and a
reduction of a deviation from a linear temperature dependence with
an increasing annealing time. (b)Dependence of resistivity on
temperature close to $T_{c}$ measured before and after annealings.
Note a systematic increase of $T_{c}$ with an increasing annealing
time.} \label{fig:fig3}
\end{figure}

\begin{figure}[ht]
\def\picfilename{Fig.44.eps}
\input epsf
\epsfxsize = 230pt \epsfysize =420pt \epsfbox{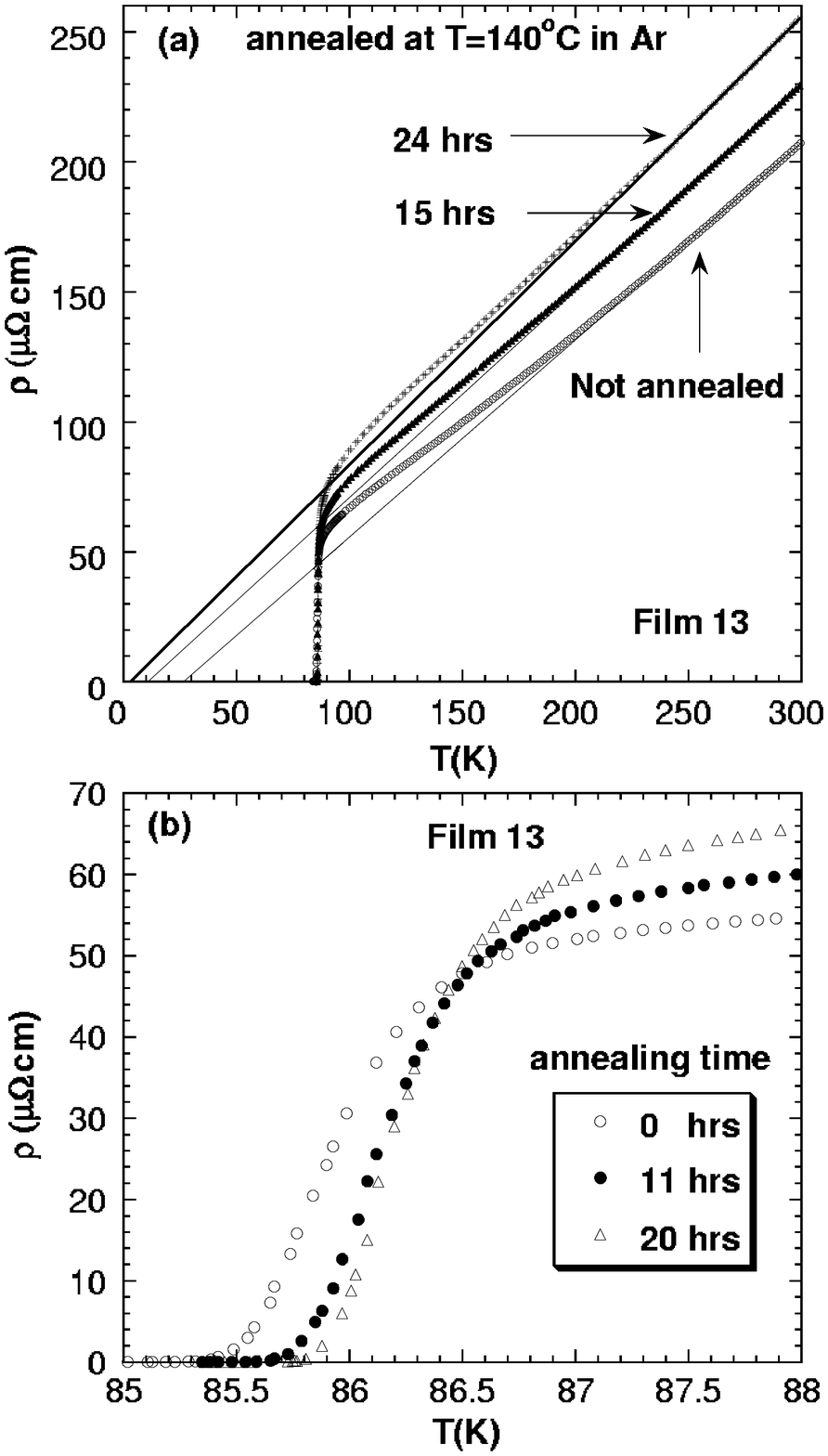}

\caption{\setlength{\baselineskip}{10pt}(a) Temperature dependence
of the in-plane resistivity $\rho_{ab}$ measured in a c-axis
oriented slightly underdoped YBCO thin film 13 with $T_{c}=85.4K$
before and after annealing for 15 and 20 hours in argon at
$140^{o}C$. Note a systematic increase of resistivity and a
reduction of a deviation from a linear temperature dependence with
an increasing annealing time. (b)Dependence of resistivity on
temperature close to $T_{c}$ before and after annealings. Note a
systematic increase of $T_{c}$ with an increasing annealing time.}
\label{fig:fig4}
\end{figure}

\begin{figure}[ht]
\def\picfilename{Fig.55.eps}
\input epsf
\epsfxsize = 230pt \epsfysize =420pt \epsfbox{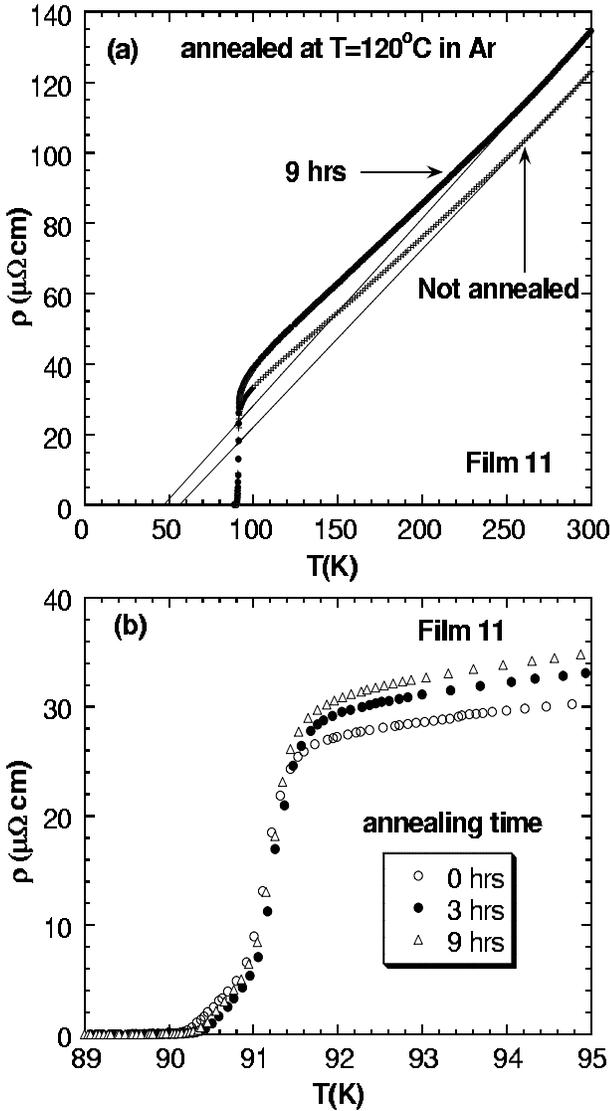}

\caption{\setlength{\baselineskip}{10pt}(a) Temperature dependence
of the in-plane resistivity $\rho_{ab}$ measured in a c-axis
oriented optimally doped YBCO thin film 11 with $T_{c}=90.4K$
before and after annealing for 9 hours in argon at $120^{o}C$. The
resistivity increases with an increasing annealing time. Deviation
from a linear temperature dependence at temperatures below
230-240K decreases with an increasing annealing time (see Fig.6
for more details). (b) Dependence of resistivity on temperature
close to $T_{c}$ measured before and after annealing. $T_{c}$ was
found to be independent of the annealing time.} \label{fig:fig5}
\end{figure}

\begin{figure}[ht]
\def\picfilename{Fig.66.eps}
\input epsf
\epsfxsize = 230pt \epsfysize =420pt \epsfbox{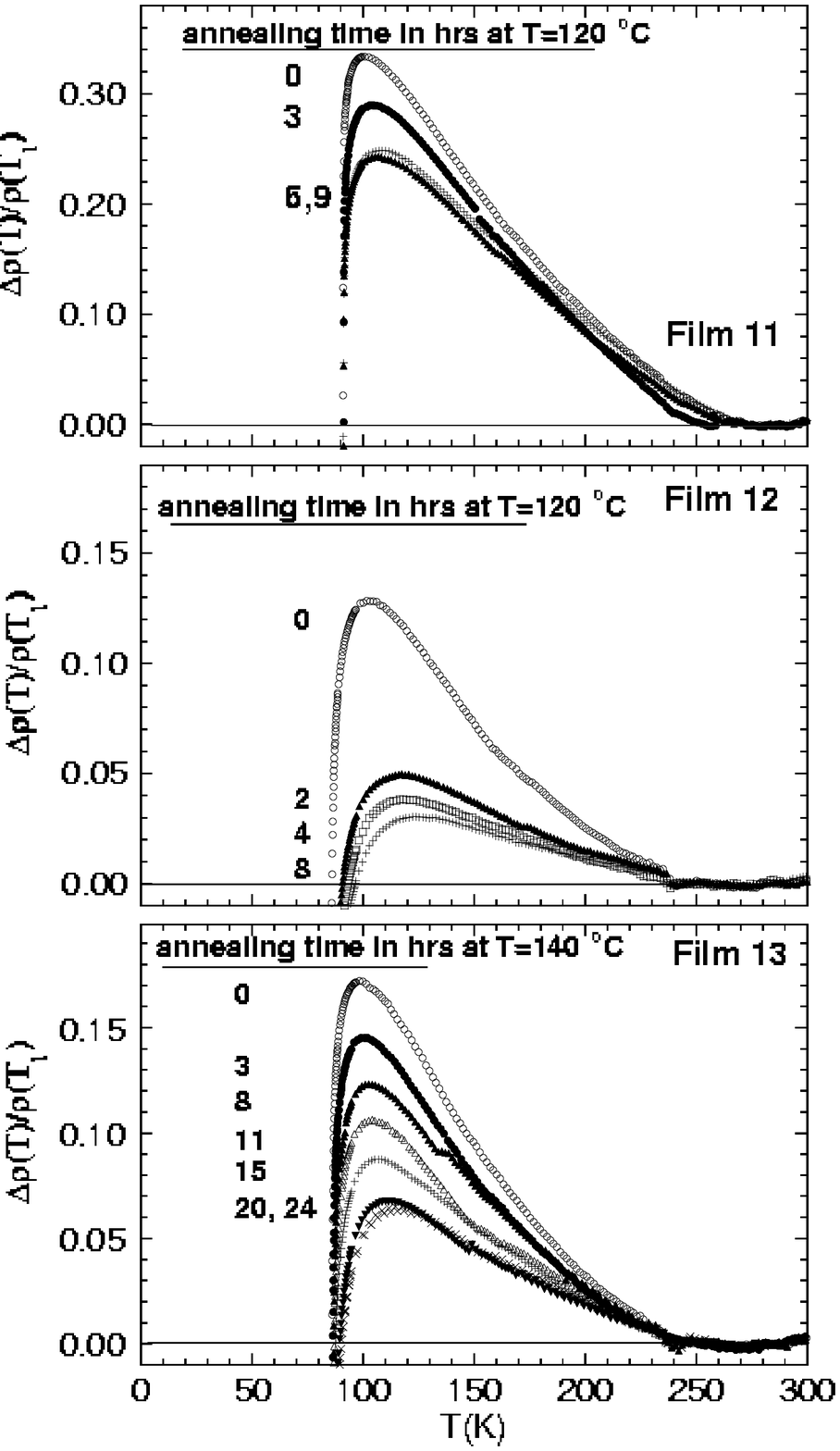}

\caption{\setlength{\baselineskip}{10pt}Upward deviation of
$\rho_{ab}(T)$ (measured for films 11,12 and 13) from a linear
temperature dependence of resistivity $\rho (T)$ defined as
$\Delta \rho(T)/\rho(T_{1})= [\rho_{ab}(T) -
\rho(T)]/\rho_{ab}(T_{1})$, where $T_{1}$ is the temperature at
which the maximum upward deviation of $\rho_{ab}(T)$ from a linear
$\rho (T)$ occurs. Note the saturation of
$\Delta\rho(T)/\rho(T_{1})$ after annealing times of 6 hours for
film 11, and 20 hours for film 13, respectively.} \label{fig:fig6}
\end{figure}

The temperature dependence of resistivity $\rho_{ab}(T)$ measured
for slightly underdoped YBCO films with $T_{c}$(R=0)= 85-86K, is
shown in Figures \ref{fig:fig3}(a) and \ref{fig:fig4}(a) as a
function of annealing time in argon at $120-140^{o}C$. The
$\rho_{ab}(T)$ for unannealed samples exhibits a flattening [an
upward deviation from a linear temperature dependence; see
straight solid line in Figures \ref{fig:fig3}(a) and
\ref{fig:fig4}(a)] at temperatures between $T_{c}$ and
approximately  220-230K. The annealing causes an increase of
$T_{c}$ [see Figures \ref{fig:fig3}(b) and \ref{fig:fig4}(b)], an
increase of resistivity, and a transformation from a non-linear
temperature dependence of resistivity towards the linear one. The
increase of $T_{c}$ depends on the sample i.e. it is about 1K
after 8 hour annealing at $120^{o}C$ for film 12, but only 0.5K
after 24 hour annealing at $140^{o}C$ for film 13. The temperature
dependence of resistivity $\rho_{ab}(T)$   measured for optimally
doped YBCO film with $T_{c}$(R=0) about 90K is shown in
Fig.\ref{fig:fig5}(a) as a function of annealing time in argon at
$120^{o}C$. The annealing of this sample leads to an increase of
the magnitude of $\rho_{ab}(T)$ and a reduction of the flattening
between $T_{c}$ and 230-240K. However, no change of $T_{c}$ was
observed in this case [see Fig.\ref{fig:fig5}(b)]. Temperature
dependence of $\rho_{ab}(T)$ at temperatures above approximately
220K is linear (see solid lines in Figures \ref{fig:fig4} and
\ref{fig:fig5}). Deviation from the linear dependence at
temperatures below 220K is shown in Fig.\ref{fig:fig6} as a
function of temperature at different annealing times. It is
defined in the following form: [$\rho_{ab}(T) -
\rho_{linear}(T)]/\rho_{ab}(T_{1})$, where $T_{1}$ is the
temperature at which a maximum upward deviation of $\rho_{ab}(T)$
from $\rho_{linear}(T)$ occurs. This deviation initially decreases
with an increasing annealing time, however it reaches saturation
at longer annealing times. The magnitude of resistivity
$\rho_{ab}(T)$ of all samples studied was found to increase
gradually with an annealing time [see Figures \ref{fig:fig4},
\ref{fig:fig5}, and \ref{fig:fig6}].

\section{DISCUSSION}

According to Jorgensen annealing of YBCO at $120-140^{o}C$ causes
redistribution of oxygen in the chain-layer, without a change in
the overall concentration of oxygen in the sample
\cite{ref:Jorgensen1}. Low temperature annealing was performed
previously by Jorgensen et al \cite{ref:Jorgensen2}and Shaked et
al \cite{ref:Shaked}. They carried out room temperature annealing
of sintered samples of underdoped YBCO (with $7-\delta=6.41$ and
6.25 ) which were quenched first  from a temperature of
$100-150^{o}C$ down to a liquid helium temperature. Room
temperature annealing led to a dramatic change of the transport
and structural properties of this compound i e, an increase of
$T_{c}$ and a decrease of lattice parameters with annealing time.
This behavior was explained in terms of a local ordering of oxygen
atoms at room temperature, which causes formation of oxygen
supercells (alternating partial and full chains) in the chain
layer. Veal et al \cite{ref:Veal} reported the observation of
time-dependent behavior at room temperature in the superconducting
and structural properties of $YBa_{2}Cu_{3}O_{x}$ single crystals,
with stoichiometries controlled within the range $6.3<x<6.6$ by
quenching the samples from $520^{o}C$ down to a liquid nitrogen
temperature. They proposed two explanations of this behavior. One
interpretation was that oxygens, quenched into O(5) sites, move to
neighboring O(1) chain sites thus enhancing the order in the O(5)
oxygen vacancy array in the chain layer. The other one suggested
that the ordering associated with low temperature annealing likely
involves the formation of alternating full and empty (oxygen free)
chains in the chain layer. Andersen et al \cite{ref:Andersen}
investigated ordering kinetics of ortho-superstructures in the
chain layers of YBCO single crystals as a function of temperature
up to about $150^{o}C$  and oxygen doping level over a range
between 6.3 and 6.8. They provided the structural phase diagram of
the superstructure ordering in YBCO (see Fig.5 in
Ref.\cite{ref:Andersen}). According to this diagram, our samples
with oxygen content of about 6.90 and 6.95 should have the ortho-I
structure with no sign of a superstructure formation between room
temperature and $150^{o}C$.

Our interpretation of the data is based on the assumption that the
redistribution of oxygen during annealing at temperatures of
$120-140 ^{o}C$ takes place in the chain layers of YBCO. It is
known that even optimally doped $RBa_{2}Cu_{3}O_{x}$ (RBCO)
(including YBCO) contains a few percent of interchain oxygen O(5)
in the chain layers \cite{ref:Jorgensen1, ref:Radaelli}.
Therefore, we also assume that O(5) oxygens form links (bridges)
between partially occupied chains of oxygen O(1). These links
could be formed by oxygen O(5) during cooling of the films from a
high deposition temperature (usually around $650-750^{o}C$). This
allows the transport current in the chain layer to zigzag between
the chains via O(5) sites, thus reducing the overall resistivity
in the chain layer. For a twinned sample, the overall resistivity
$\rho_{ab}$ is an average of $\rho_{a}$ and $\rho_{b}$ in the a-
and b-direction, respectively ie., $\rho_{ab} = (\rho_{a} +
\rho_{b})/2$. It is also equal to the plane- and the
chain-layer-resistivities connected in parallel according to the
formula,

\begin{equation}\label{eq:Rab1}
\frac{1}{\rho_{ab}}=\frac{1}{\rho_{plane}}+\frac{1}{\rho_{chain-layer}}
\end{equation}
or

\begin{equation}\label{eq:Rab2}
\rho_{ab} = \frac{\rho_{plane}}{1
+\frac{\rho_{plane}}{\rho_{chain-layer}}}
\end{equation}

According to the experimental results obtained on untwinned YBCO
crystals\cite{ref:Rice,ref:Ito,ref:Gagnon}, temperature dependence
of resistivity in the planes is linear i e.,

\begin{equation}\label{eq:Rplane}
\rho_{plane}(T) = a_{1} + a_{2}T
\end{equation}

where $a_{1}$ and $a_{2}$ are constants. On the other hand,
according to references \cite{ref:Rice,ref:Ito}, temperature
dependence of the chain resistivity in untwinned YBCO crystals
contains both linear and quadratic components. The latest
reference\cite{ref:Gagnon}, revealed however that on average the
chain resistivity of four untwinned YBCO crystals follows almost
pure $T^{2}$ dependence over a temperature range between 130 and
300 K, i e:

\begin{equation}\label{eq:Rchain}
\rho_{chain}(T) = b_{1} + b_{2}T^{2}
\end{equation}

where $b_{1}$ and $b_{2}$ are constants. In equations
Eq.\ref{eq:Rplane} and Eq.\ref{eq:Rchain} resistivity and
temperature are in units of $\mu\Omega cm$ and K, respectively.

$T^{2}$ dependence of resistivity was observed in quasi-1D organic
conductors (see for example Weger et al. \cite{ref:Weger}).
Abrikosov and Ryzhkin \cite{ref:Abrikosov} developed a theory of
the temperature dependence of resistivity in quasi-1D metals.
According to this theory in quasi-1D metals with a long
mean-free-path, electron-phonon scattering produces $T^{2}$
dependence of resistivity. Such model may be applied to the chains
of YBCO, since in pure YBCO the mean-free-path should be much
larger than the lattice constant. So far however no calculations
of $\rho_{chain}(T)$ have been performed using the relevant phonon
frequencies for YBCO.

\begin{figure}[ht]
\def\picfilename{Fig.7m.eps}
\input epsf
\epsfxsize = 200pt \epsfysize =250pt \epsfbox{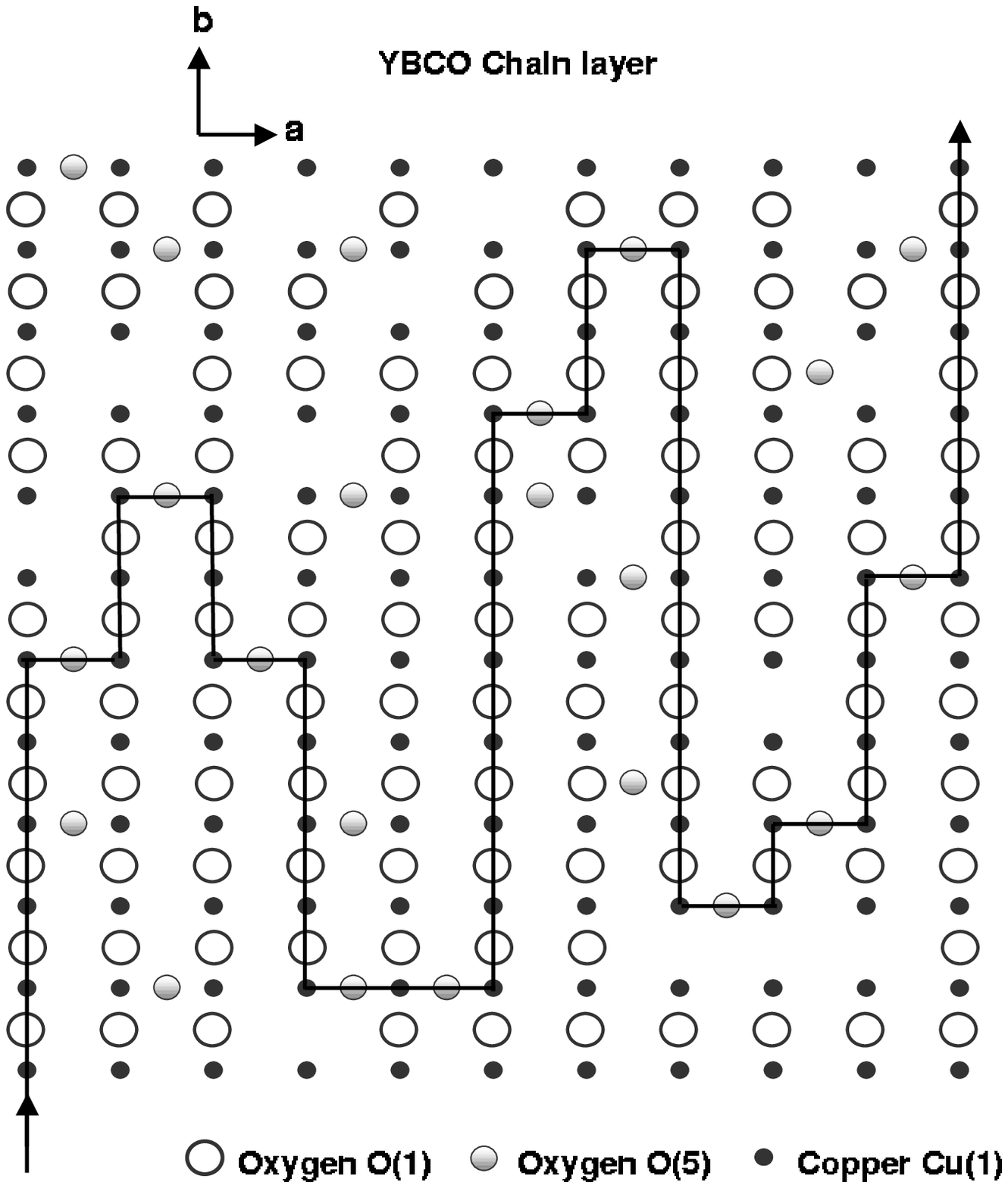}

\caption{\setlength{\baselineskip}{10pt} Representation of the
proposed local structure of the chain layer. Chain oxygen O(1)
ions are represented by open circles,interchain oxygen O(5) by
shaded circles, and copper ions by black dots. Black solid line
represents possible zigzag path of the transport current through
O(5) sites in the chain layer.} \label{fig:fig7}
\end{figure}

In an underdoped YBCO, and even in an optimally doped YBCO, the
chains are not fully occupied. As mentioned above, in the presence
of interchain O(5), the transport current could therefore zigzag
between the chains in the chain layer via O(5) bridges along a
path of the lowest resistance (see Fig.\ref{fig:fig7}). In this
case electrical transport would still be quasi-1D and the
resistivity should be proportional to $T^{2}$. One also could
expect that O(5) bridges lower the overall resistance in the chain
layer. The activation energy for the motion of O(5) oxygen between
chains in the b-direction is very small \cite{ref:Rothman} and
during annealing at relatively low temperatures of $120-140^{o}C$
oxygen O(5) should be highly mobile in this direction. High
mobility of O(5) during annealing could change the electrical
properties of the chain layer. O(5) can fill some O(1) vacancies
in the chains, which leads to a higher $T_{c}$. On the other hand,
the motion of O(5) between the chains in the chain direction could
interrupt the paths of the lowest resistance, causing an overall
increase of resistivity in the chain layer. One could make a
counter-argument that even in the absence of interchain O(5)
oxygen, annealing could cause more disorder in the chains,
resulting in an increase of resistivity. This however should be
accompanied by a decrease in $T_{c}$.

We considered the following model that could explain an increase
of resistivity with the annealing time in twinned YBCO thin films.
In the presence of oxygen O(5) the electrical transport in the
chain layer is determined by the resistance of the chain fragments
and the interchain bridges through O(5) sites. The resistance of
the interchain bridges R between any two neighboring chains could
be considered as that of average N "effective" bridges (through
which the current flows) connected in parallel ie. $R = R_{o}/N$,
where $R_{o}$ is the resistance of a single effective bridge.
During annealing at $120-140^{o}C$ in argon, due to high mobility
of oxygen O(5) the number of effective bridges could be reduced.
We assumed that this process is completely random and the number
of effective bridge disintegrations that occurs, is proportional
to the number of effective bridges present ie.,

 \begin{equation}\label{eq:dN/dt}
\frac{dN}{dt} = - \alpha N
\end{equation}

where $\alpha$ is a decay constant. Dependence of N on time is
then:

\begin{equation}\label{eq:N(t)}
N = N_{0} \exp{(-\alpha t)}
\end{equation}

where t is the annealing time and $N_{o}$ is the number of
effective bridges at t=0. The resistance R should then increase
with time t according to the expression:

\begin{equation}\label{eq:R(t)}
R = R_{0}/N = (R_{0}/N_{0}) \exp{(\alpha t)}
\end{equation}

The total average resistance offered to the current by any two
chains could be written as the sum of the resistance due to chain
fragments (which we consider time independent) and that due to the
time dependent O(5) bridges [Eq.(\ref{eq:R(t)})]. Therefore, the
overall resistivity in the chain layer as a function of annealing
time t and temperature T can be expressed in the following form,

  \begin{eqnarray}\label{eq:rho}
    \rho_{chain-layer}(t,T) &=& \rho_{0}(T)\frac{1 + C \exp{(\alpha t)}}{1 + C} \\
    &=& \frac{b_{1} + b_{2}T^{2}}{1 + C}(1 + C \exp{(\alpha t)})
    \nonumber
  \end{eqnarray}


where $\rho_{0}(T) = b_{1} + b_{2}T^{2}$ is the total chain layer
resistivity at temperature T and annealing time t=0 (which
includes contributions due to both chain fragments and O(5)
bridges), and C is the normalization constant which ensures that
at t=0 the total chain layer resistivity equals $\rho_{0}(T)$.

Then the overall resistivity $\rho_{ab}(t,T)$, including the
contribution due to the planes, could be written as

\begin{equation}\label{eq:rhoab}
\rho_{ab}(t,T) = \frac{\rho_{plane}(T)}{1 +
\frac{\rho_{planes}(T)}{\rho_{chain-layer}(t,T)}}
\end{equation}

We assumed that $\rho_{plane}(T)$ in this equation depends on
temperature according to Eq.(\ref{eq:Rplane}) and not on the
annealing time, so all changes in resistivity due to annealing
occur in the chain layers.

\begin{figure}[ht]
\def\picfilename{Fig.77n.eps}
\input epsf
\epsfxsize = 230pt \epsfysize =420pt \epsfbox{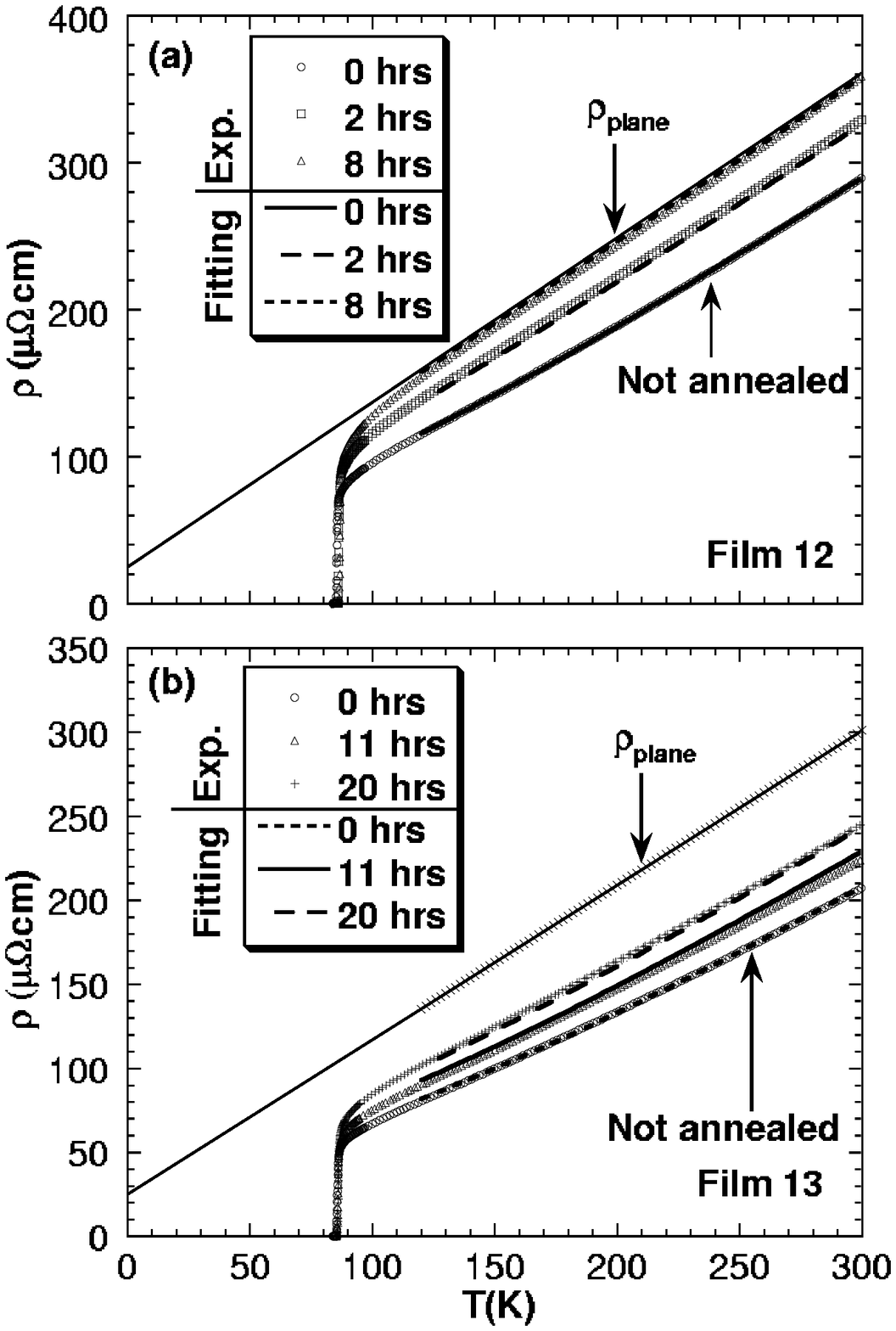}

\caption{\setlength{\baselineskip}{10pt} Experimental data and
theoretical fitting curves obtained between 120 and 300K, using
the model described in the text, for $\rho_{ab}(t,T)$ measured in
slightly underdoped films 12 and 13. The assumed resistivity of
the planes $\rho_{planes}(T)$ in Eq.(\ref{eq:rhoab}) is
represented by the solid line marked $\rho_{plane}$ at
temperatures between 120 and 300K.}
 \label{fig:fig8}
\end{figure}

\begin{figure}[ht]
\def\picfilename{Fig.88n.eps}
\input epsf
\epsfxsize = 230pt \epsfysize =220pt \epsfbox{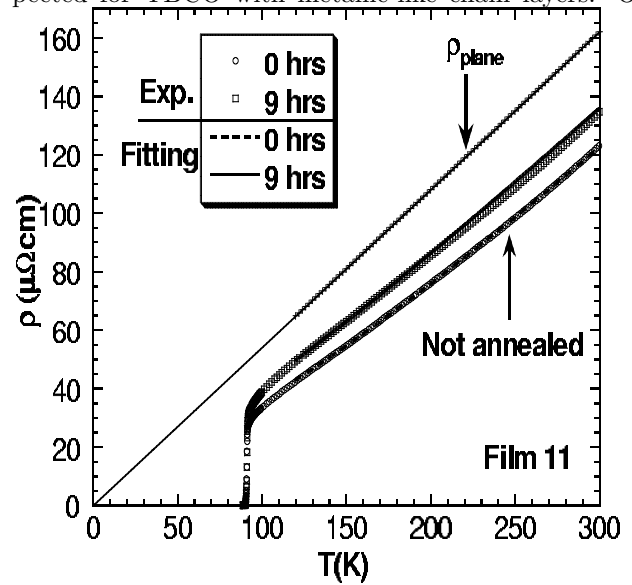}

\caption{\setlength{\baselineskip}{10pt} Experimental data and
theoretical fitting curves obtained between 120 and 300K, using
the model described in the text, for $\rho_{ab}(t,T)$ measured in
an optimally doped film 11. The assumed resistivity of the planes
$\rho_{planes}(T)$ in Eq.(\ref{eq:rhoab}) is represented by the
solid line marked $\rho_{plane}$ at temperatures between 120 and
300K.} \label{fig:fig9}
\end{figure}

We performed computer calculations of $\rho_{ab}(t,T)$ and
compared them with the data obtained for twinned  c-axis oriented
YBCO films. The fits to the experimental data are shown in Figures
\ref{fig:fig8} and \ref{fig:fig9}. The fitting parameters are
$a_{1}$, $\alpha$, and C. Constants $a_{1}$ and $a_{2}$ in the
equation $\rho_{plane} (T) = a_{1} + a_{2}T$ are shown in Table
\ref{tab:tab1}  for films 11, 12, and 13 and for
 untwinned YBCO crystals. The slopes of
$\rho_{plane}$ curves in Figures \ref{fig:fig8} and
\ref{fig:fig9}, which are represented by constants $a_{2}$, were
assumed to follow roughly the slope of the experimental data for
$\rho_{ab}(t,T)$ at temperatures above 250K. Constants $b_{1}$ and
$b_{2}$ for $\rho_{chain}(T)$ (see Eq.(\ref{eq:Rchain})) were
calculated using Eq.(\ref{eq:Rab2}), Eq.(\ref{eq:Rplane}) and the
experimental data for $\rho_{ab}(t=0,T)$ in unannealed (t=0) YBCO
films. Good agreement between the experimental data and the model
was obtained for both optimally doped and slightly underdoped YBCO
films.

\begin{table}
 \centering \caption{The parameters $a_{1}$, $a_{2}$, $\alpha$,
and C, which were used in the calculation of $\rho_{ab}(t,T)$ for
YBCO films 11,12 and 13. For untwinned YBCO crystals, the
experimental values of $a_{1}$ and $a_{2}$ are shown. Note:
$a_{1}$ and $a_{2}$ for untwinned YBCO crystals 1-4
\cite{ref:Gagnon} represent the average values for all four
crystals. $a_{1}$, $\alpha$, and C are the fitting parameters.}
\vspace{5 mm}

\begin{tabular}{c c c c c c}
\hline
 Material & $a_{1}$ & $a_{2}$ & $\alpha$ & C & Ref \\
 \hline\hline
YBCO film 11 & 0 & 0.54 & 0.066 & 5 & this work \\
YBCO film 12 & 25 & 1.12 & 0.442 & 10 & this work  \\
YBCO film 13 & 25 & 0.92 & 0.042 & 2 & this work \\
YBCO untwinned crystal 1 & 16.5 & 1.00 &  &  &\cite{ref:Gagnon} \\
YBCO untwinned crystals 1-4 & 0 & 1.00 &  &  &\cite{ref:Gagnon} \\
YBCO untwinned crystal & -10.5 & 0.79 &  &  & \cite{ref:Rice} \\
YBCO untwinned crystal & -10.3 & 0.76 &  &  &\cite{ref:Ito} \\

\hline
\end{tabular}
\label{tab:tab1}
\end{table}

According to Eq.(\ref{eq:rhoab}) the total resistivity
$\rho_{ab}(t,T)$ approaches that of the planes (which increases
linearly with an increasing temperature) if the chain layer
resistivity is very high (i.e. it exhibits a semi-insulating or
insulating behavior). This could happen after a long annealing
time t if the decay constant $\alpha$ (which describes the process
of disintegration of O(5) bridges between O(1) chains), is
independent of time. On the other hand, if the chain layer
resistivity is small (i.e. a metallic-like), then the chain layer
contributes substantially to the total resistivity. In this case,
one could expect a deviation of $\rho_{ab}(t,T)$ from a linear
temperature dependence towards a quadratic-like one, due to a
$T^{2}$ dependence of $\rho_{chain}(T)$. Eq.(\ref{eq:rhoab}) shows
that the change of resistivity from a small towards a large one in
the chain layer (which corresponds to a transition from a
metal-like to an insulating behavior), causes an increase of the
total resistivity $\rho_{ab}(T)$ and a transformation of its
temperature dependence from a non-linear towards a linear one.
This could explain why so many YBCO single crystal samples (or
YBCO thin film samples) with the same $T_{c}$ have very different
magnitudes and temperature dependencies of resistivity.

\section{CONCLUSIONS}
Our investigation of the temperature dependence of resistivity of
twinned YBCO thin films as a function of annealing time at $
120-140^{o}C$ led us to conclude that the resistive properties of
the chain layer (located between two $CuO_{2}$ double-planes)
could have a dramatic effect on the magnitude and the temperature
dependence of the total resistivity $\rho_{ab}(T)$. We suggest
that the resistive properties of the chain layer could be modified
by the residual amount of the interchain oxygen O(5) which allows
a current to flow between the chains along a path of the lowest
resistance. Following this suggestion, annealing at $120-140^{o}C$
causes a  redistribution of oxygen O(5) and a resulting disruption
of low resistance paths and an increase of the chain-layer
resistivity. Distribution of oxygen O(5) in the chain layer could
therefore be responsible for the change of the resistive state of
this layer from a metallic-like state to an insulating one.
Halbritter\cite{ref:Halbritter} argued that the resistivity of the
chain layer in YBCO affects the resistivity along c-axis and
consequently the anisotropy of the normal state resistivity. Small
conductivity in the chain layer should therefore cause higher
anisotropy (and higher c-axis resistivity) than that expected for
YBCO with metallic-like chain layers. Our recent measurements of
the angular dependence of resistivity in the ab-planes close to
the superconducting transition temperature revealed larger than
expected changes of resistivity in some YBCO thin films, when the
angle between the direction of an external magnetic field and the
ab-planes was changed from 0 to 90 degrees \cite{ref:Abdelhadi}.
These large changes suggest large anisotropy (and strong pinning
of magnetic flux between the $CuO_{2}$ planes), which could be
caused by an insulating state of the chain layers in these
samples. On the other hand, the changes of chain-layer resistivity
may be also responsible for different temperature dependencies of
c-axis resistivity observed in different YBCO single crystals.
Surprisingly, the c-axis resistivity of YBCO crystals studied so
far exhibits either a $1/T$ dependence \cite{ref:Penney,ref:Hagen}
or a linear temperature dependence
\cite{ref:Iye,ref:Ossipyan,ref:Friedmann}.

\section{Acknowledgements}
 \vspace{3mm} This work was supported  by a grant
from the Natural Sciences and Engineering Research Council of
Canada (NSERC). We benefited from useful discussions with Z.W.
Gortel. We are grateful to M. Denhoff, R. Hughes and J.R. Preston
for supplying us with YBCO thin films.




\end{document}